\documentclass[twocolumn]{aastex6}
\usepackage{amsmath}

\begin{document}

\title{Stellar Companions to the Exoplanet Host Stars HD 2638 and HD
  164509}

\author{
  Justin M. Wittrock\altaffilmark{1},
  Stephen R. Kane\altaffilmark{1},
  Elliott P. Horch\altaffilmark{2},
  Lea Hirsch\altaffilmark{3},
  Steve B. Howell\altaffilmark{4},
  David R. Ciardi\altaffilmark{5},
  Mark E. Everett\altaffilmark{6},
  Johanna K. Teske\altaffilmark{7}
}
\email{jwittroc@mail.sfsu.edu}
\altaffiltext{1}{Department of Physics \& Astronomy, San Francisco
  State University, 1600 Holloway Avenue, San Francisco, CA 94132,
  USA}
\altaffiltext{2}{Department of Physics, Southern Connecticut State
  University, New Haven, CT 06515, USA}
\altaffiltext{3}{Astronomy Department, University of California at
  Berkeley, Berkeley, CA 94720, USA}
\altaffiltext{4}{NASA Ames Research Center, Moffett Field, CA 94035,
  USA}
\altaffiltext{5}{NASA Exoplanet Science Institute, Caltech, MS 100-22,
  770 South Wilson Avenue, Pasadena, CA 91125, USA}
\altaffiltext{6}{National Optical Astronomy Observatory, 950 N. Cherry
  Ave, Tucson, AZ 85719, USA}
\altaffiltext{7}{Carnegie Department of Terrestrial Magnetism, 5241
  Broad Branch Road, NW, Washington, DC 20015, USA}


\begin{abstract}

An important aspect of searching for exoplanets is understanding the
binarity of the host stars. It is particularly important because
nearly half of the solar-like stars within our own Milky Way are part
of binary or multiple systems. Moreover, the presence of two or more
stars within a system can place further constraints on planetary
formation, evolution, and orbital dynamics. As part of our survey of
almost a hundred host stars, we obtained images at 692~nm and 880~nm
bands using the Differential Speckle Survey Instrument (DSSI) at the
Gemini-North Observatory. From our survey, we detect stellar
companions to HD~2638 and HD~164509. The stellar companion to HD~2638
has been previously detected, but the companion to HD~164509 is a
newly discovered companion. The angular separation for HD~2638 is
$0.512 \pm 0.002\arcsec$ and for HD~164509 is $0.697 \pm
0.002\arcsec$.  This corresponds to a projected separation of $25.6
\pm 1.9$ AU and $36.5 \pm 1.9$ AU, respectively.  By employing stellar
isochrone models, we estimate the mass of the stellar companions of
HD~2638 and HD~164509 to be $0.483 \pm 0.007$ $M_\sun$ and $0.416 \pm
0.007$ $M_\sun$, respectively, and their effective temperatures to be
$3570 \pm 8$~K and $3450 \pm 7$~K, respectively. These results are
consistent with the detected companions being late-type M dwarfs.

\end{abstract}

\keywords{planetary systems -- techniques: high angular resolution --
  stars: individual (HD~2638, HD~164509)}


\section{Introduction}

Much of the focus in the exoplanetary field still lies in the
detection of planets using a variety of techniques, such as
radial velocity (RV) signatures, transits, direct imaging,
microlensing, among others. A significant factor that can affect the
detection of exoplanets is the binarity of the host stars. In fact, it is
believed that nearly half of all sun-like stars are part of a multiple-star
system \citep{raghavan10}. This high-rate of multiplicity has also been
found in exoplanet host stars through follow-up of {\it
  Kepler} candidates \citep{eve15,kra16} and Robo-AO observations of
RV exoplanet host stars \citep{riddle15}.

The mere presence of a binary companion can substantially affect
astrometric and RV measurements of the host star, and cause severe
blended contamination for transit experiments
\citep{car15,cia15,gil15}. It is therefore imperative to verify the
multiplicity of exoplanet host stars to ensure correct interpretation
of exoplanet signals. Moreover, the binarity of the stars can place
further constraints on planetary formation. \citet{holman99} explored
the orbital stability of the planets in the presence of a binary star
system. Additionally, correlations between planets' mass and their
period \citep{zucker02} and eccentricities \citep{eggenberger04} were
examined. Several binary systems have been studied, such as $\alpha$
Centauri \citep{benest88}, Sirius \citep{benest89}, $\eta$ Coronae
Borealis \citep{benest96}, and 30 Arietis B \citep{kane15,roberts15},
which provide us rich information on orbital dynamics in a N-body system.

This paper presents new results on stellar companions to the exoplanet
host stars HD~2638 and HD~164509. The stellar companion to HD~2638 has
been previously detected and characterized \citep{riddle15,roberts15}.
However, this is an independent detection, and this paper shall present
independent analysis of that system.  In the meanwhile, the companion to
HD~164509 has not been previously reported. In Section~\ref{properties} we
briefly describe the properties of HD~2638 and HD~164509, along with their
known exoplanets. Section~\ref{obs} discusses the method of detection, the
range of targets that were selected for analysis, and the details of
the data reduction. Section \ref{res} presents the results from the
data analysis and stellar isochrone fitting. Section \ref{impl}
explains the potential implication of those findings for the planetary
systems, including limits to the eccentricities of the binary
companion that allow orbital stability. Section \ref{con} provides
discussion of further work and concluding remarks.

\begin{deluxetable*}{lCC}
  \tablecolumns{3}
  \tablewidth{0pt}
  \tablecaption{\label{prop} Stellar \& Planetary Properties}
  \tablehead{
    Properties &
    \colhead{HD~2638\tablenotemark{a, b}} &
    \colhead{HD~164509\tablenotemark{c}}}
  \startdata
  \sidehead{Stellar}
  ~~~~Spectral Type\tablenotemark{d}									&	G5V					&	G5V				\\
  ~~~~$M_{\star}$ ($M_{\sun}$)\tablenotemark{e}						&	0.87 \pm 0.03		&	1.10 \pm 0.01	\\
  ~~~~$R_{\star}$ ($R_{\sun}$)\tablenotemark{e}						&	0.81 \pm 0.02		&	1.11 \pm 0.02	\\
  ~~~~$L_{\star}$ ($L_{\sun}$)\tablenotemark{e}						&	0.42 \pm 0.01		&	1.31 \pm 0.02	\\
  ~~~~$T_e$ (K)\tablenotemark{e}										&	5173 \pm 26 		&	5860 \pm 31		\\
  ~~~~$\log$ $g$ ($cm/s^2$)\tablenotemark{e}							&	4.55 \pm 0.03		&	4.38 \pm 0.02	\\
  ~~~~Age (Gyr)\tablenotemark{e}										&	5.1 \pm 4.1			&	3.2 \pm 0.8		\\
  ~~~~$[$Fe/H$]$														&	0.16 \pm 0.05 		&	0.21 \pm 0.03	\\
  ~~~~Apparent Magnitude m$_V$\tablenotemark{f}						&	9.58				&	8.24			\\
  ~~~~Proper Motion ($\alpha$, $\delta$) (mas/yr)\tablenotemark{f}	&	-105.63, -223.46	&	-7.40, -20.98	\\
  ~~~~Parallax (mas)\tablenotemark{f}									&	20.03 \pm 1.49		&	19.07 \pm 0.97	\\
  ~~~~Distance (pc)\tablenotemark{f}									&	49.93 \pm 3.71		&	52.44 \pm 2.67	\\
  \sidehead{Planetary}
  ~~~~$M_p \sin i$ ($M_J$)	&	0.48						&	0.48 \pm 0.09	\\
  ~~~~P (Days)				&	3.43752 \pm 0.00823876		&	282.4 \pm 3.8	\\
  ~~~~a (AU)					&	0.044						&	0.875 \pm 0.008 \\
  \enddata
  \tablenotetext{a}{\citet{wang11}}
  \tablenotetext{b}{\citet{moutou05}}
  \tablenotetext{c}{\citet{giguere12}}
  \tablenotetext{d}{\citet{esa97}}
  \tablenotetext{e}{\citet{bonfanti16}}
  \tablenotetext{f}{\citet{leeuwen07}}
\end{deluxetable*}


\section{Properties of the HD~2638 and HD~164509 Systems}
\label{properties}

The detailed stellar and planetary parameters of the HD~2638 and
HD~164509 systems are shown in Table \ref{prop}. HD~2638 is a G5V star
that is about 50~pc away toward the constellation of Cetus
\citep{esa97, leeuwen07}. It is believed to be part of a wide binary
system with the nearby star HD~2567. \citet{shaya11} performed a
Bayesian analysis of both stars' astrometry; the result yielded 99\%
chance of both stars being true companions. However, \citet{roberts15}
argued that, barring any errors in the measurement of the stars'
parallax, they are separated by 6.8 pc, making them not
gravitationally bound. HD~2638 is known to host one planet, HD 2638b,
with a mass of approximately 0.48 $M_J$
\citep{moutou05}. \citet{riddle15} discovered that HD~2638 has a
stellar companion while examining the system with ROBO-AO.
\citet{roberts15} analyzed the orbital dynamics of the primary star
and the stellar companion and determined that the
masses of the components are 0.87 $M_{\sun}$ and 0.46
$M_{\sun}$, respectively.  Moreover, they inferred that the spectral
types are G8V and M1V and that they are separated by about
28.5 AU, giving them an orbital period of around 130 years
\citep{roberts15}. \citet{ginski16} performed additional astrometric
and photometric analysis on the system and found that the mass of the
companion star is $0.425^{+0.067}_{-0.095}$ $M_{\sun}$.

HD~164509 is a G5V star that is about 52~pc away toward the
constellation of Ophiuchus \citep{esa97, leeuwen07}. It is known to
host one planet, HD 164509b, with a mass of approximately 0.48 $M_J$
\citep{giguere12}.  \citet{giguere12}, upon examining the RV data,
found that it displays ``a residual linear trend of $−5.1 \pm 0.7$ m
s$^{-1}$ year$^{-1}$, indicating the presence of an additional longer
period companion in the system''. \citet{sirothia14} studied this
system and reported a 150 MHz radio signature of $18 \pm 6$ mJy. The
authors speculated that it could be the cause of a massive moon
``orbiting a rapidly-rotating giant planet''; however, they emphasized
that more analysis is needed before such a conclusion can be reached.


\section{Observations and Data Reduction}
\label{obs}

Speckle observations of our target stars were obtained with the
Differential Speckle Survey Instrument, or DSSI \citep{horch09}. This
instrument was built at Southern Connecticut State University by one
of us (E. H.), and currently enjoys official visitor instrument status
at the Gemini-North Observatory. The observations were carried out as
part of a larger survey program that aims to detect low-mass stellar
companions to exoplanet host stars. Observations were carried out in
July, 2014 when $\sim$60 targets were observed. Each measurement was
acquired using two different passbands, one at 692~nm and another at
880~nm. The 692~nm filter has FWHM of 40~nm, and the 880~nm filter has
FWHM of 50~nm. After all images underwent data reduction, they were
directly examined using the ds9 program for any bright source
appearing next to the target. The two particular targets described
here, HD~164509 and HD~2638, were observed during the night of 2014
July 22 and 23 respectively. The results from the remainder for the
survey targets will be published elsewhere.

Final reconstructed images were produced from the speckle data
sequences using methods that have been described in previous papers
\citep[e.g.][]{horch12, horch15}, but we will briefly describe the
main points here. The raw speckle data are stored as FITS data cubes
consisting of 1000 frames, where each frame is a $256 \times
256$-pixel image centered on the target. Frames are bias-subtracted,
and then an autocorrelation is formed. These are then summed to
generate a final autocorrelation for the entire observation. We
Fourier transform this to obtain the spatial frequency power spectrum
of the observation. The same operations are then performed on an
unresolved star (effectively a point source) that lies close on the
sky to the science target. By dividing the power spectrum of the
science target by that of the point source, we deconvolve the effects
of the speckle statistics, and arrive at a diffraction-limited
estimate of the true power spectrum of the object.

Returning to the raw data frames, we next form the image bispectrum of
each frame, which is the Fourier transform of the triple correlation,
as described in \citet{lohmann83}. This data product is known to
contain information that can be used to calculate the phase of the
object's Fourier transform, which we do using the relaxation algorithm
of \citet{meng90}. By taking the square root of the deconvolved power
spectrum and combining it with this phase estimate, we generate a
diffraction-limited estimate of the (complex) Fourier transform of the
object. Finally we multiply this with a Gaussian low-pass filter of
width similar to the diffraction limit of the telescope, and
inverse-transform to arrive at the final reconstructed image.

\begin{figure*}
  \begin{center}
    \begin{tabular}{cc}
      \includegraphics[width=8cm]{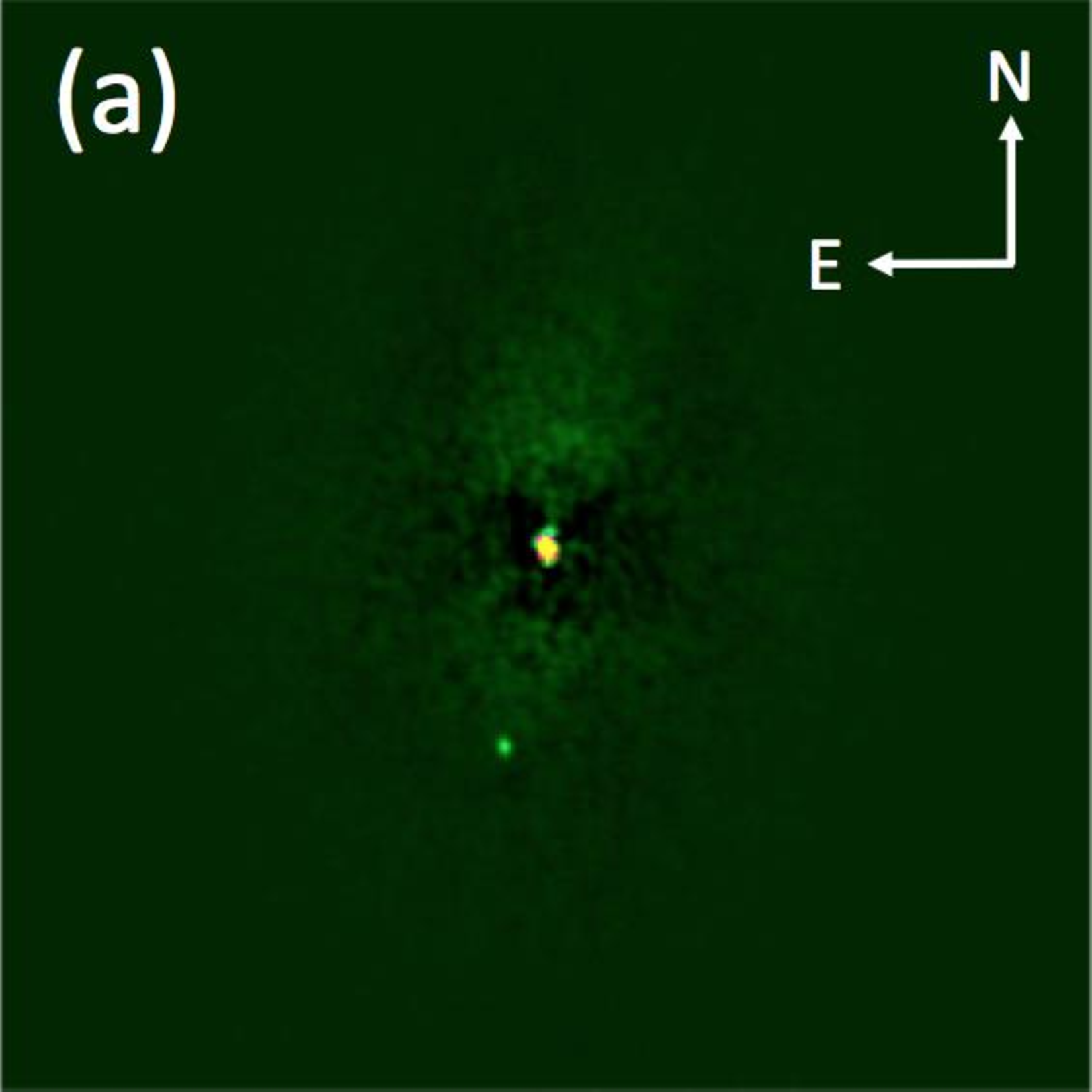} &
      \includegraphics[width=8cm]{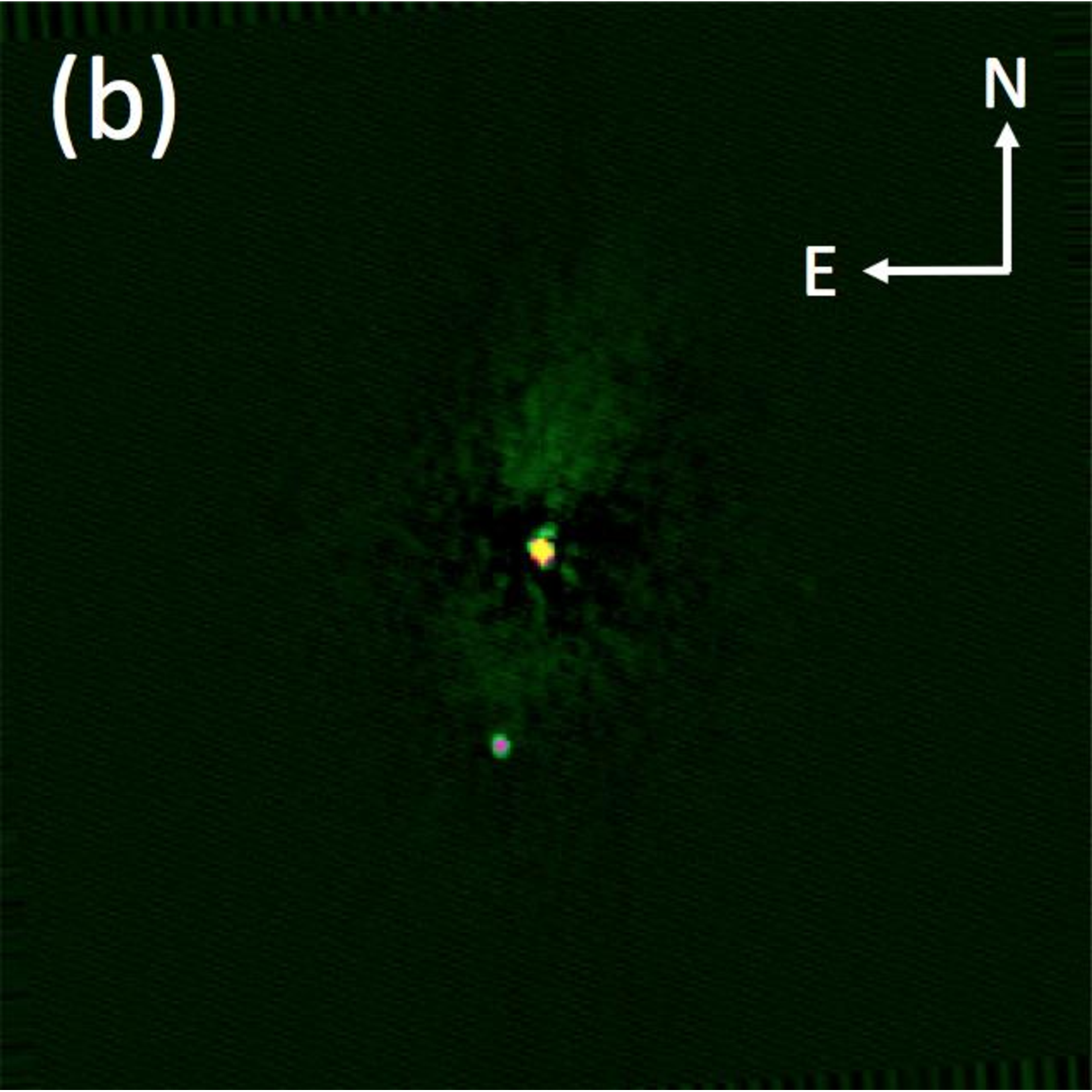} \\
      \includegraphics[angle=90,width=8cm]{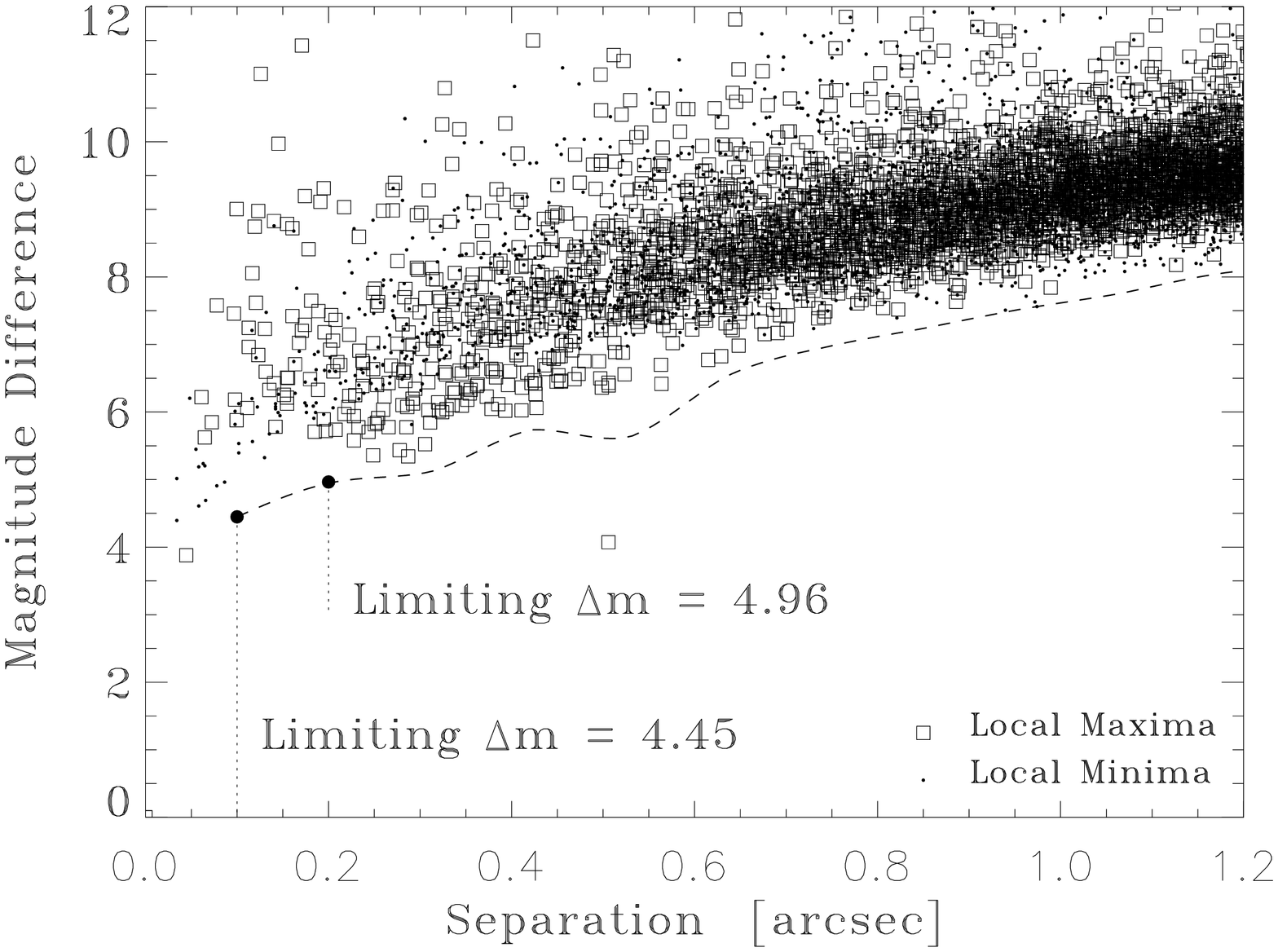} &
      \includegraphics[angle=90,width=8cm]{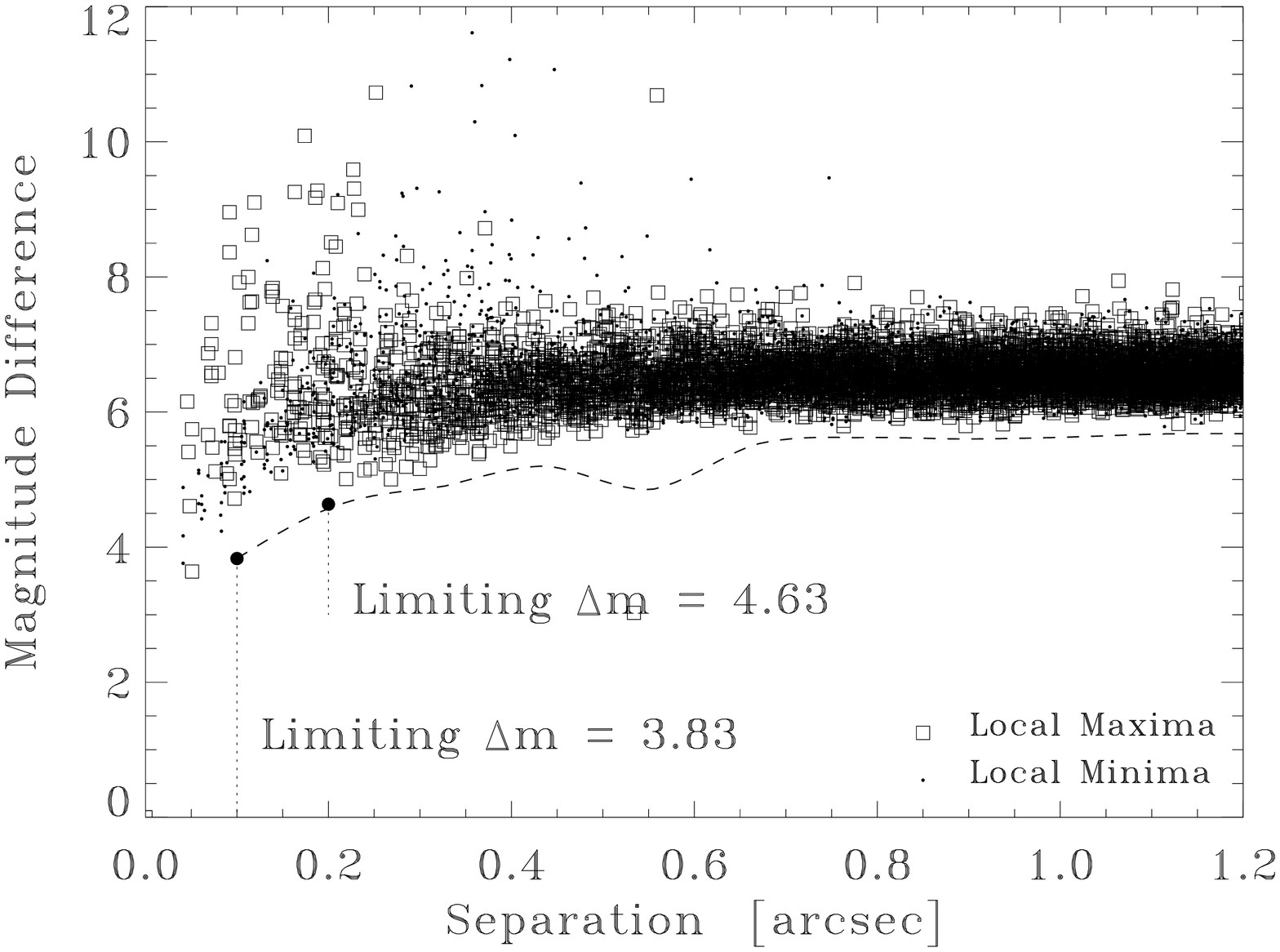}
    \end{tabular}
  \end{center}
  \caption{Top left and top right are Gemini DSSI speckle images of
    HD~2638 at 692~nm and 880~nm, respectively. The field-of-view is
    $2.8 \times 2.8$''. As indicated in both images, North is up and
    East is to the left.  The source in the center is HD~2638, and a
    bright source to the bottom and slightly to the left of the main
    star is the stellar companion. Botton left and bottom right are
    sensitivity plots of HD~2638 at 692~nm and 880~nm,
    respectively. Each plot shows the limiting magnitude (difference
    between local maxima and minima) as a function of apparent
    separation from HD~2638 in arcsec. The dashed line is a cubic
    spline interpolation of the 5$\sigma$ detection limit. Both plots
    were generated from top left and right corresponding images.}
  \label{hd2638}
\end{figure*}

\begin{figure*}
  \begin{center}
    \begin{tabular}{cc}
      \includegraphics[width=8cm]{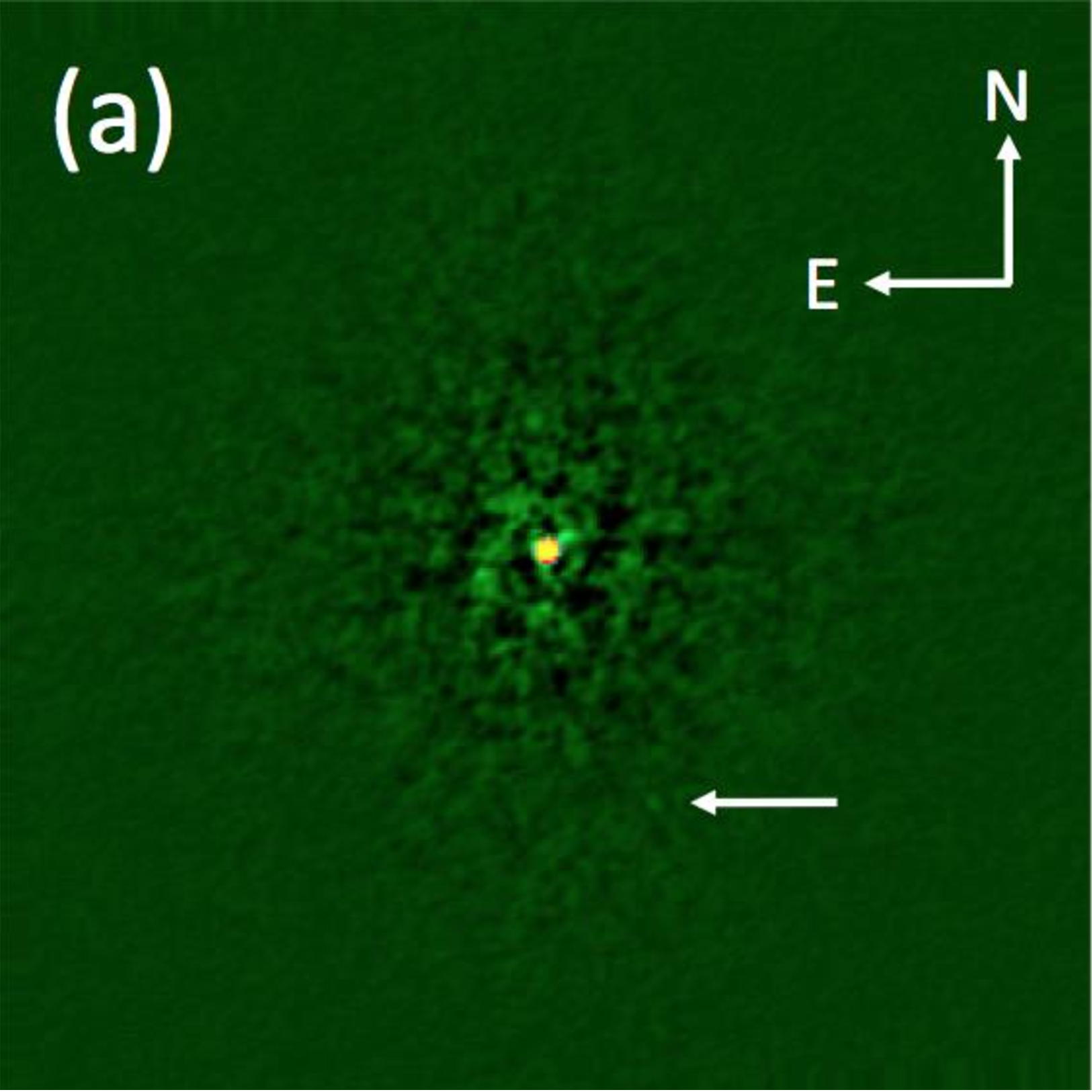} &
      \includegraphics[width=8cm]{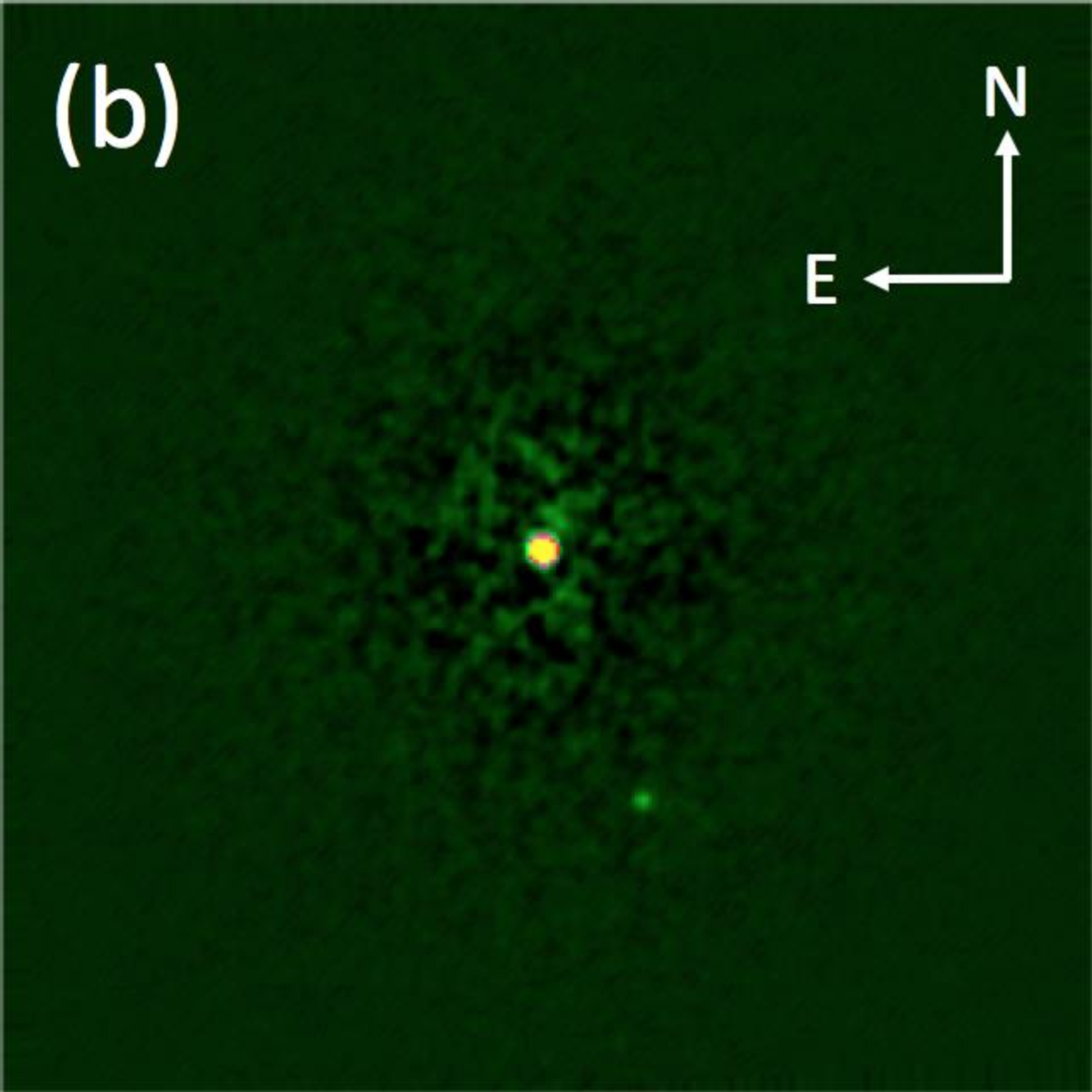} \\
      \includegraphics[angle=90,width=8cm]{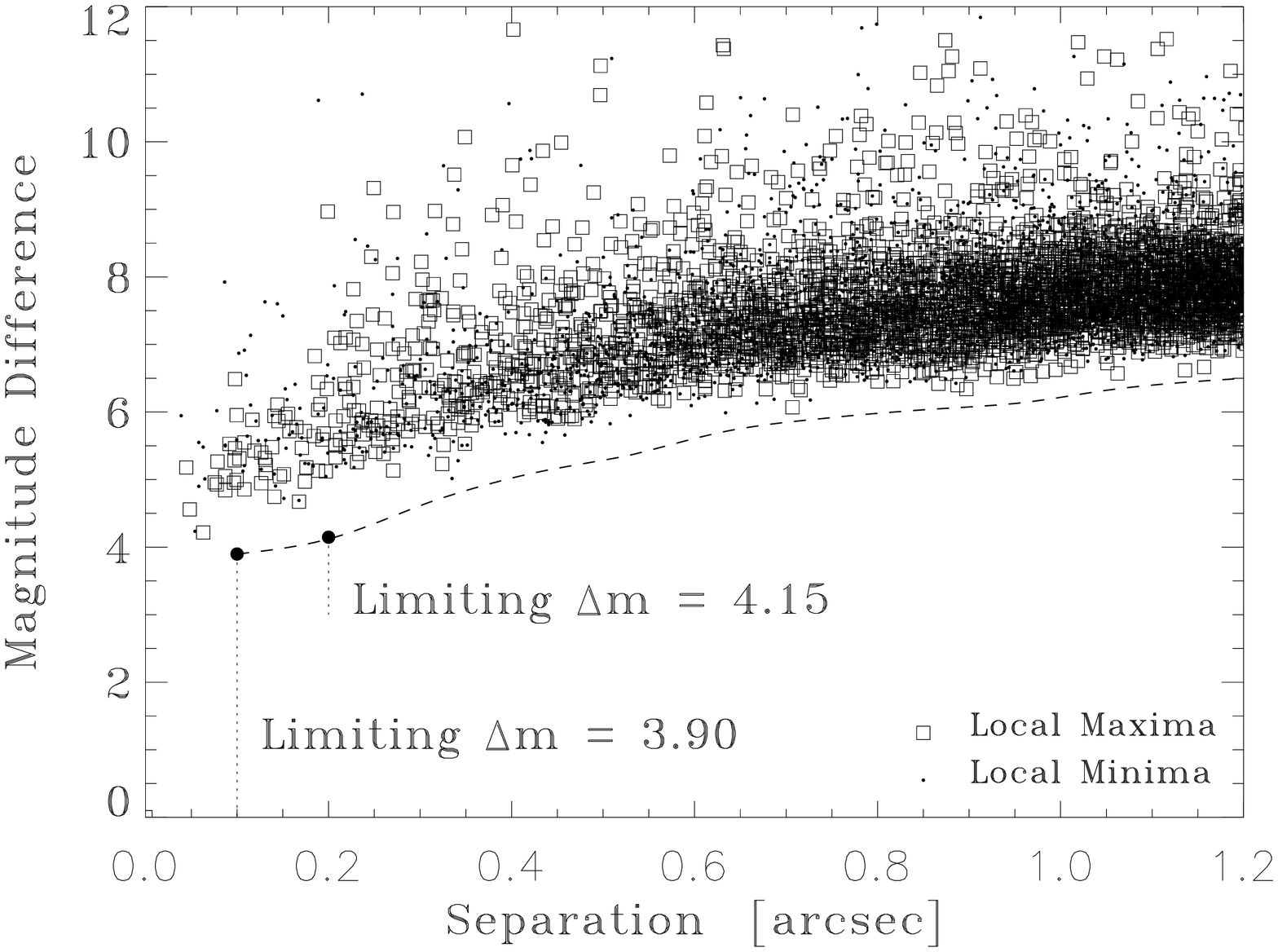} &
      \includegraphics[angle=90,width=8cm]{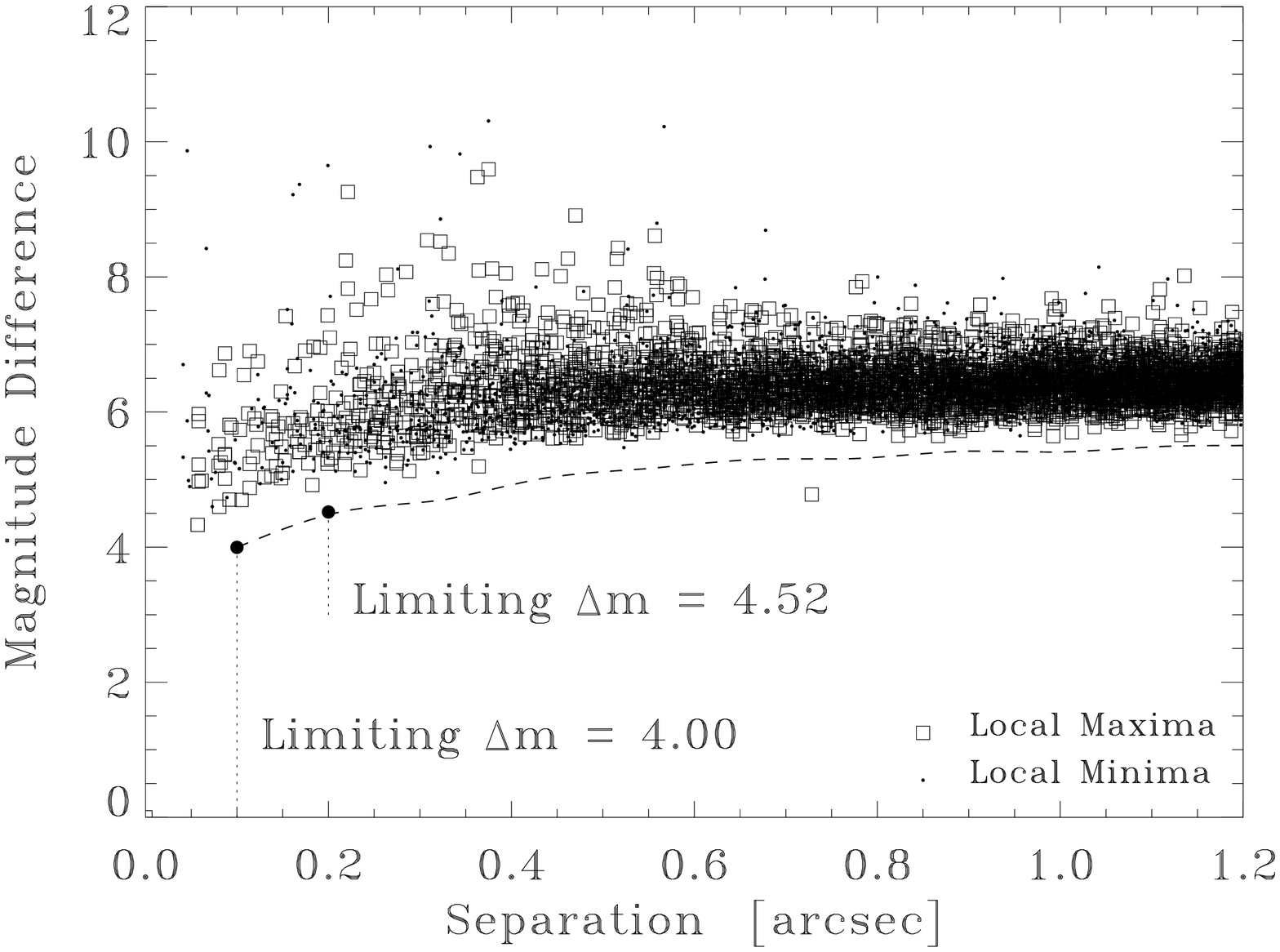}
    \end{tabular}
  \end{center}
  \caption{Top left and top right are Gemini DSSI speckle images of HD
    164509 at 692 nm and 880~nm. The field-of-view is $2.8 \times
    2.8$''. As indicated in both images, North is up and East is to
    the left. The source in the center is HD~164509, and a bright
    source to the bottom and right of the main star is the stellar
    companion. The arrow in the left image indicates the location of
    the companion. Bottom left and bottom right are sensitivity plots
    of HD~164509 at 692~nm and 880~nm, respectively. Each plot shows
    the limiting magnitude (difference between local maxima and
    minima) as a function of apparent separation from HD~164509 in
    arcsec. The dashed line is a cubic spline interpolation of the
    5$\sigma$ detection limit. Both plots are generated from top left
    and right corresponding images.}
  \label{hd164509}
\end{figure*}

Using the reconstructed images, we can study the statistics of local
maxima that occur as a function of separation from the central star in
order to derive a detection limit curve versus separation. We follow
the method described in \citet{horch11}. By computing the average and
standard deviation of the maxima inside annuli that have different
mean separations from the primary star, we estimate the 5-sigma
detection limit as the mean value plus five times the standard
deviation, converted to a magnitude difference. For Gemini data, this
is done centering annuli at distances of 0.1, 0.2, 0.3, ..., 1.2
arcsec. We then use a cubic spline interpolation to develop a smooth
detection limit curve at all separations in between the two extreme
limits. Curves like this are shown in Figures \ref{hd2638} and
\ref{hd164509}.


\section{Results}
\label{res}

Target stars that were imaged using DSSI have been examined; however,
only the images of two stars, HD 2638 and HD 164509, show a nearby
bright source, as can be seen in both Figures \ref{hd2638} and
\ref{hd164509}. Results of the DSSI observations for both stars are
tabulated in Table \ref{data}. As described by \citet{horch15},
typical uncertainties in separation for DSSI at Gemini are 1--2
milli-arcsecs (mas). For a particularly faint component, such as the
companion to HD~164509, the uncertainty will lie at the upper end of
that range. Thus, for separation, we assigned the conservative
uncertainty of 2~mas, as shown in Table~\ref{data}. Similarly for the
position angle, an uncertainty of $\sim$$0.2\degr$ is consistent with
previous measurements acquired using the Gemini/DSSI
configuration. For uncertainties in the magnitude difference between
primary and secondary, we used the empirically determined precision
for such measurements provided by \citet{horch04}.

The sensitivity plots provided in Figures \ref{hd2638} and
\ref{hd164509} show the magnitude difference between local maxima and
minima in the corresponding image as a function of the separation from
the primary host star. The construction of these sensitivity plots are
described in more detail by \citet{how11}. As shown in the sensitivity
plots, the limiting resolution of DSSI with Gemini is
$\sim$$0.05\arcsec$. Each of the two stars discussed here and their
corresponding results are described separately below.


\subsection{HD 2638}

Prior to submitting this work, we learned that \citet{riddle15} has
detected HD~2638's stellar companion. We present our results as an
independent detection of this companion. Both DSSI images from Figure
\ref{hd2638} show a bright source to the bottom and slightly to the
left of HD~2638. Based on the magnitude differences from Table
\ref{data}, the stellar companion appears to be brighter at 880~nm
than it is at 692~nm, implying that the stellar companion is a
late-type star. According to \citet{roberts15}, the companion's
spectral type is M1V, which seems to be in agreement with our
assessment. Our calculations of the projected separation between
HD~2638 and its companion star yield $25.5 \pm 1.9$ AU at 692~nm and
$25.6 \pm 1.9$ AU at 880~nm, which is close to \citet{roberts15}'s
28.5 AU physical separation. Note that the apparent close companion to
the north of the primary in each image is within the limting
resolution of the instrument and is thus an artifact of the speckle
image processing.


\subsection{HD 164509}

Figure \ref{hd164509} contains two images that display a source
southwest of HD~164509. The magnitude differences of this system imply
that the stellar companion is considerably fainter than the host star
by a factor of almost 100. In fact, it is so faint at 692 nm that it
is difficult to resolve in the image. Despite the fact that HD~164509
is more luminous than HD~2638, the considerable faintness of
HD~164509's companion as compared to HD~2638's implies that this
stellar companion is a very cool, late-type star. Based on the data
from Table \ref{data}, the physical separation between HD~164509 and
its companion is $36.5 \pm 1.9$~AU. To compare, the planet Neptune is
about 30~AU from our Sun, and the result falls short of the dwarf
planet Pluto's average distance of 39.5~AU. Since HD~164509 is
slightly more massive than our Sun and with the given distance between
the host star and its companion, this leads to credence that the faint
object may be gravitationally bound to HD~164509. One interesting
thing to point out is that this dim star may be ``an additional longer
period companion" that \citet{giguere12} speculated when they came
across the RV data's residual linear trend. As of this writing, there
has been no confirmation of HD~164509 hosting a stellar companion.

\begin{deluxetable*}{lCCCC}
  \tablecolumns{5}
  \tablewidth{0pt}
  \tablecaption{\label{data} DSSI Astrometry \& Photometry Results}
  \tablehead{
    Measurements &
    \multicolumn{2}{c}{HD~2638} &
    \multicolumn{2}{c}{HD~164509} \\
    \colhead{} &
    \colhead{692 nm} &
    \colhead{880 nm} &
    \colhead{692 nm} &
    \colhead{880 nm}}
  \startdata
  Position Angle E of N ($\degr$)	&	167.7 \pm 0.2		&	167.7 \pm 0.2		&	202.5 \pm 0.2		&	202.6 \pm 0.2	\\
  Apparent Separation ($\arcsec$)	&	0.511 \pm 0.002		&	0.513 \pm 0.002		&	0.697 \pm 0.002		&	0.697 \pm 0.002	\\
  Projected Separation (AU)		&	25.5 \pm 1.9		&	25.6 \pm 1.9		&	36.5 \pm 1.9		&	36.5 \pm 1.9	\\
  $\Delta$m$^{*}$					&	3.83 \pm 0.2		&	2.80 \pm 0.2		&	5.53 \pm 0.4		&	4.41 \pm 0.4
  \enddata
  \tablenotetext{*}{Note: $\Delta$m is the apparent magnitude
    difference between the primary and secondary stars.}
\end{deluxetable*}


\subsection{Stellar Isochrone Fitting}

To determine the properties of the detected stellar companions, we
performed a stellar isochrone fit using the methodology described by
\citet{eve15} and \citet{teske15}. Briefly, the method maps out the
probability distribution of the primary star using Dartmouth stellar
isochrones. The inputs for this analysis are the stellar properties
shown in Table~\ref{prop}. The combination of the resulting
probability distributions for the primary with the multi-band
observations described in Section~\ref{obs} produce a probability
distribution for the properties of the secondary. Such a result
assumes that it is a bound companion that falls on the same isochrone
as the primary.

\begin{deluxetable}{lCC}
  \tablecolumns{3}
  \tablewidth{0pt}
  \tablecaption{\label{isotable} Stellar Companion Isochrone Fitting Results}
  \tablehead{
    Parameters &
    \colhead{HD~2638} &
    \colhead{HD~164509}}
  \startdata
	~~~~$M_{\star}$ ($M_{\sun}$) & 0.48 \pm 0.03 & 0.42 \pm 0.03 \\
	~~~~$R_{\star}$ ($R_{\sun}$) & 0.46 \pm 0.02 & 0.40 \pm 0.02 \\
	~~~~$L_{\star}$ ($L_{\sun}$) & 0.030 \pm 0.005 & 0.020 \pm 0.003 \\
	~~~~$T_e$ ($K$)              & 3571 \pm 48      & 3446 \pm 43 \\
	~~~~$\log$ $g$ ($cm/s^2$)    & 4.80 \pm 0.02 & 4.85 \pm 0.02
  \enddata
\end{deluxetable}

\begin{figure*}
  \begin{center}
    \begin{tabular}{cc}
      \includegraphics[width=8cm]{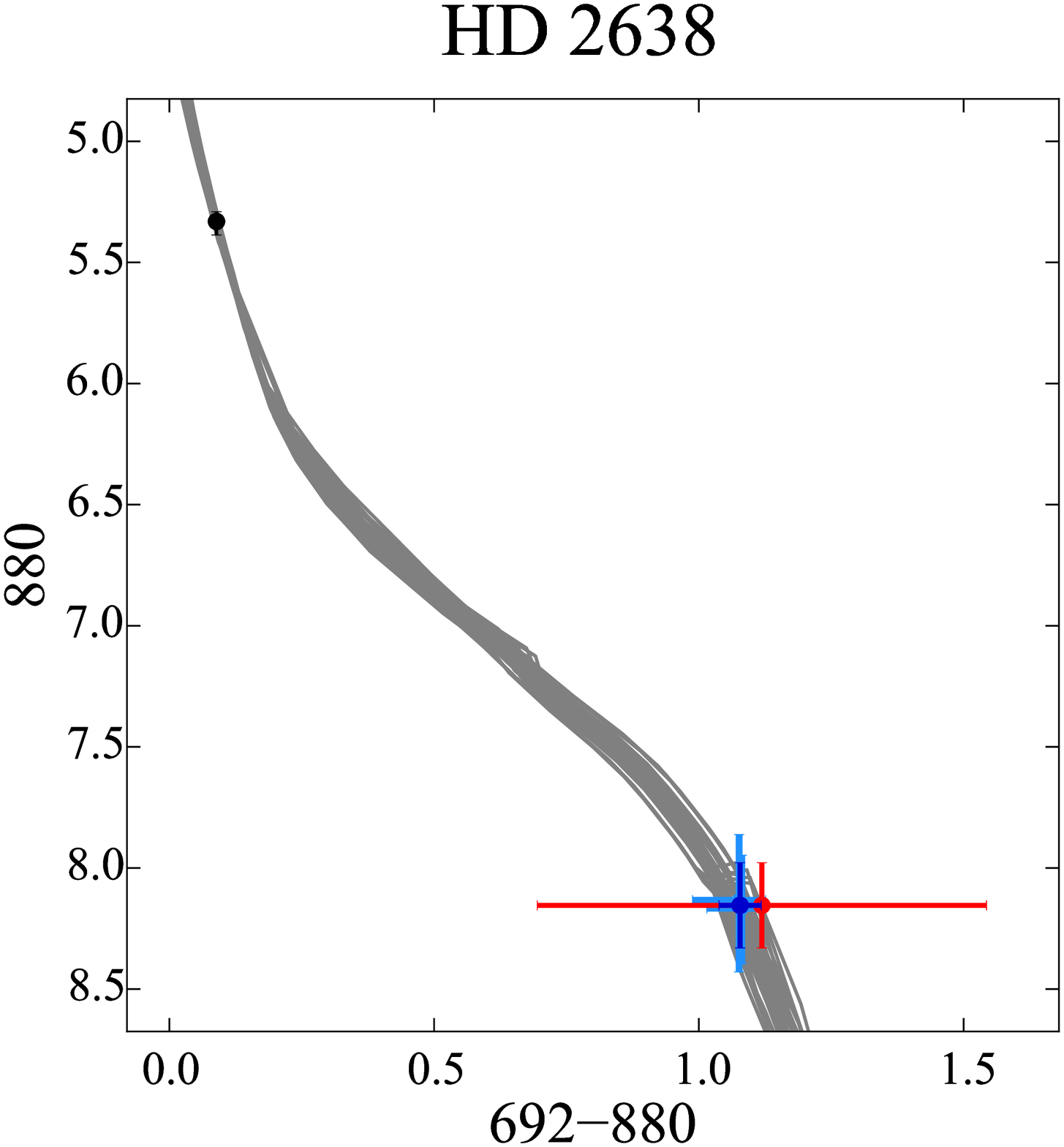} &
      \includegraphics[width=8cm]{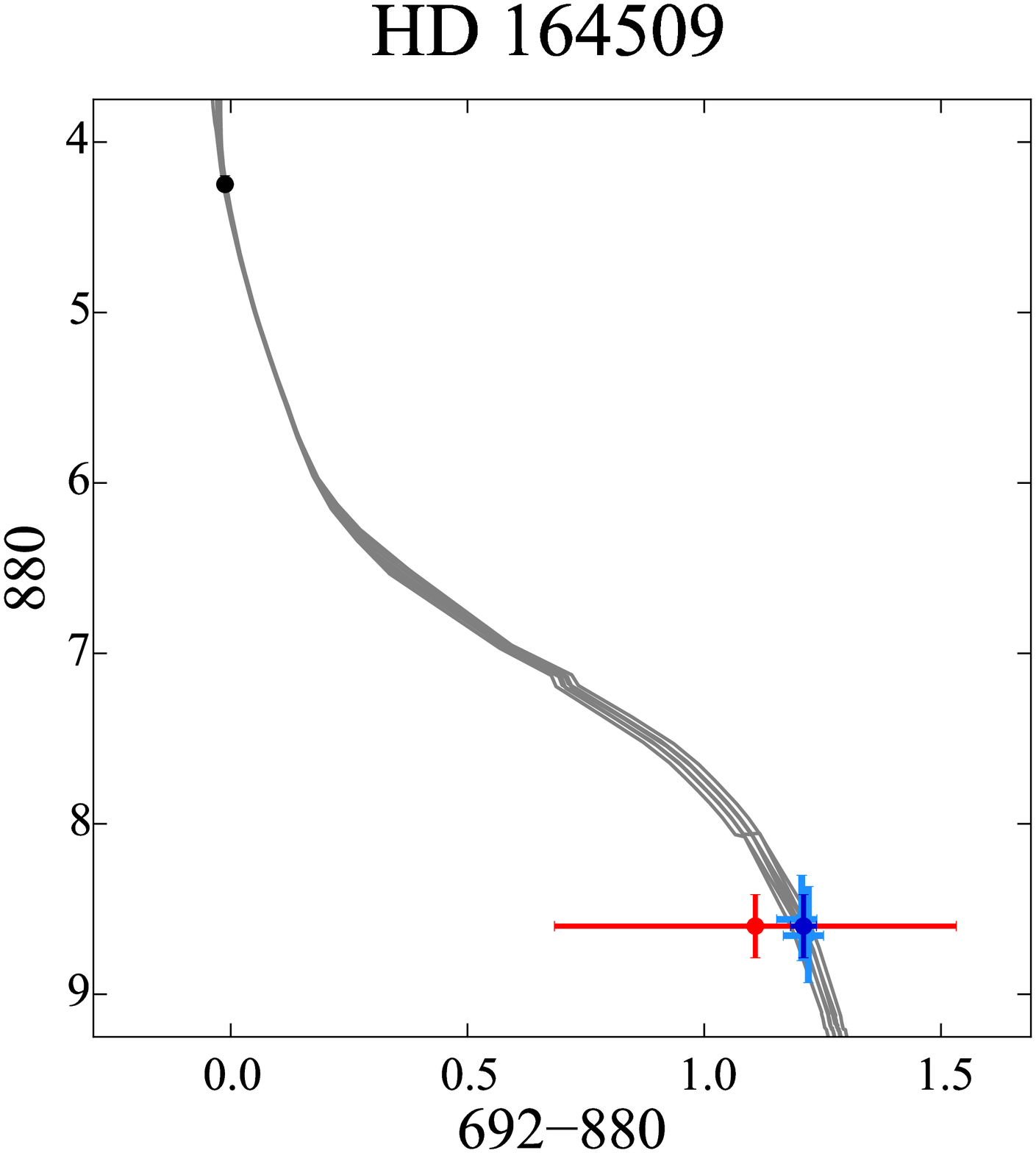}
    \end{tabular}
  \end{center}
  \caption{Stellar isochrone models of HD~2638 (left) and HD~164509
    (right). The black dot near the top is the primary star, the dark
    blue dot near the bottom is the average model of the companion,
    and the red dot is the observed companion. Note that in both
    cases, the model fits well with the observation.}
  \label{isoplot}
\end{figure*}

The results of our isochrone fits for the stellar companions are shown
in Table \ref{isotable}. The derived stellar properties are consistent
with both of the companion stars being late-type main sequence stars
(M dwarfs). Note that both our mass determination
(Table~\ref{isotable}) and projected separation (Table~\ref{data}) for
the HD~2638 stellar companion match well with the results obtained by
\citet{roberts15}. The results from our isochrone fitting are shown in
Figure~\ref{isoplot} for HD~2638 (left) and HD~164509 (right). The
color-magnitude diagrams include the set of isochrones that are within
$\pm 1 \sigma$ of the primary star metallicity. The black data point
represents the primary star and the red data point shows the location
of the secondary based upon the measurements described in
Section~\ref{obs}. The dark blue data point is the average location of
the secondary based upon the probability distributions of the
isochrone fitting. The location of the secondary from measurements and
from isochrone fits are consistent with one another, indicating that
the assumption of the secondary being bound to the primary is indeed a
valid assumption.


\subsection{Proper Motion and Astrometry}

\begin{figure}
  \includegraphics[width=8.2cm]{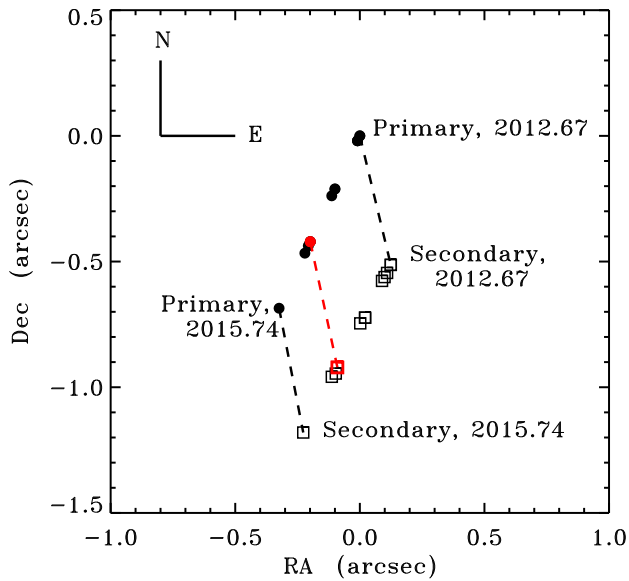}
  \caption{The proper motion of the HD~2638 primary (solid circles)
    and secondary (open squares) over time. The measurements presented
    in this work are shown in red. The dashed lines link the primary
    and secondary for the first and last observations in the sequence,
    and for our observation.}
  \label{astrofit}
\end{figure}

A further test that the detected companions are indeed bound to the
primary is to analyse the common proper motion of the stars on the
sky. For HD~2638, such data are available from the ``Fourth
Interferometric Catalog of Binary Stars'' (see description in
\citet{har01}). The observations of HD~2638 from the catalog span a
time frame from 2012.67 to 2015.74. Figure~\ref{astrofit} shows the
locations of the primary star and secondary star relative to a zero
position for the primary at the first epoch shown, 2012.67. The
primary star positions are shown as filled circles, and the secondary
positions as open squares. The two measurements presented in this
work, which on this scale are indistinguishable, are shown in
red. Dotted lines link the primary and secondary for the first and
last observations in the sequence, and for our 692-nm
observations. The proper motions (see Table~\ref{prop}) are drawn from
\citet{leeuwen07}. This figure demonstrates that the pair of stars are
clearly moving together.

For HD~164509, we have only the single measurement described in this
work for the relative astrometry as of now, so the same analysis
cannot be completd. However, it is worth noting that the proper motion
of HD~164509 is significantly smaller than for HD~2638 (see
Table~\ref{prop}). Thus, a few more speckle observations of this star
over the next few years would allow the same analysis to be undertaken
since these numbers, although not as big as for HD~2638, are several
times the typical precision for the speckle observations.


\section{Implications for the Known Planets}
\label{impl}

\begin{figure*}
  \begin{center}
    \begin{tabular}{cc}
      \includegraphics[width=8cm]{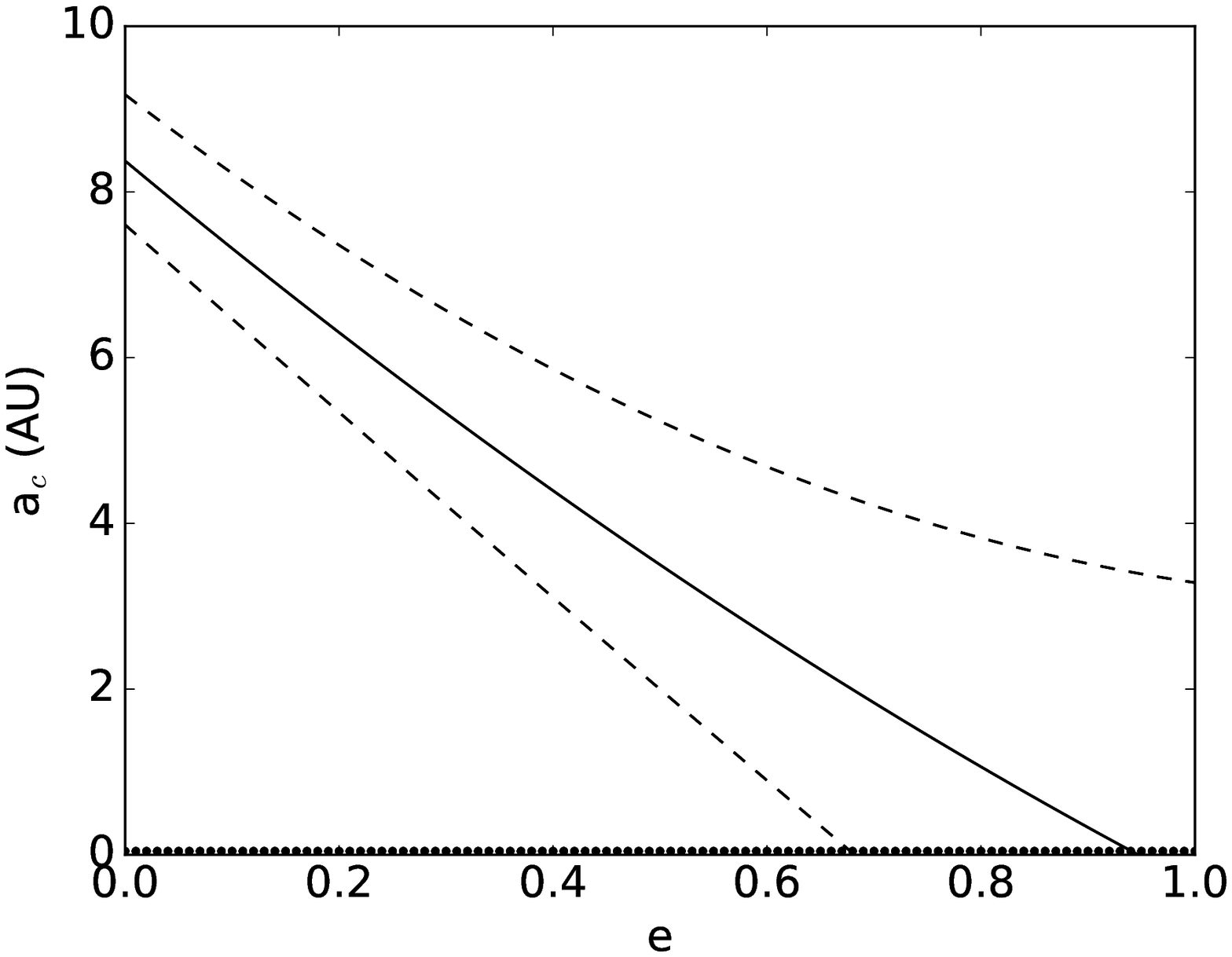} &
      \includegraphics[width=8cm]{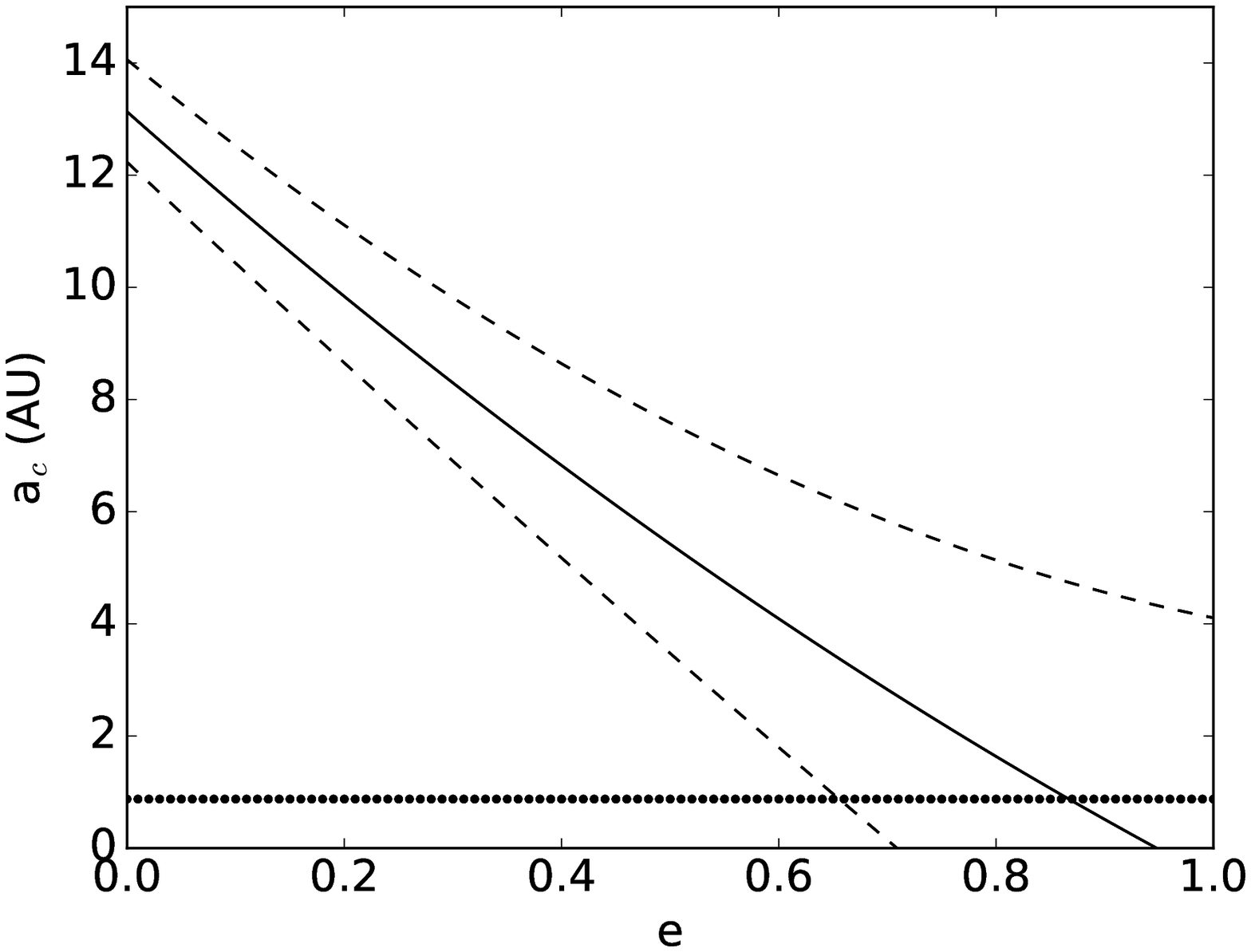}
    \end{tabular}
  \end{center}
  \caption{Plots of critical semi-major axis $a_c$ vs orbital
    eccentricity $e$ (solid line) for HD~2638 (left) and HD~164509
    (right). The dashed lines indicate the 1$\sigma$ uncertainties in
    the relationship and the horizontal dotted lines represent the
    semi-major axes of the known planets.}
  \label{ecclimits}
\end{figure*}

The presence of stellar companions can pose a significant challenge
for orbital stability and formation scenarios for planets in such
systems \citep{ngo15,wang15}. Issues regarding planet formation
include protoplanetary disc truncation, grain condensation, and
planetesimal accumulation (see \citet{the14} and references
therein). To test that our observations are consistent with the
presence of the planets, we use the orbital dynamics results of
\citet{holman99} for test particles in binary systems. Specifically,
we calculate the critical semi-major axis, $a_c$, beyond which
planetary orbits would be unstable in the systems. The resulting plots
of $a_c$ as a function of binary eccentricity $e$ are shown in
Figure~\ref{ecclimits}. The values of $a_c$ were calculated using
Equation 1 from \citet{holman99}:
\begin{equation}
  \begin{split}
    a_c = & [(0.464 \pm 0.006) + (-0.380 \pm 0.010) \mu \\
    & + (-0.631 \pm 0.034)e + (0.586 \pm 0.061) \mu e \\
    & + (0.150 \pm 0.041) e^2 \\
    & + (-0.198 \pm 0.074) \mu e^2] a_b
  \end{split}
  \label{eqnac}
\end{equation}
where $a_b$ is the binary semi-major axis.  The mass ratio, $\mu$, is
calculated as $\mu = m_2 / (m_1 + m_2)$ where $m_1$ and $m_2$ are the
masses of the primary and secondary respectively. Using the values
from Tables \ref{data} and \ref{isotable}, we have $\mu = 0.357 \pm
0.009$ \& $a_b = 25.5$~AU for HD~2638, and $\mu = 0.274 \pm 0.006$ \&
$a_b = 36.5$~AU for HD~164509. Note that this assumes the projected
separations are the true semi-major axes of the binary
companions. Including both the uncertainties in Equation~\ref{eqnac}
and $\mu$, we include lines for the 1$\sigma$ uncertainties as dashed
lines in Figure~\ref{ecclimits}. The semi-major axes of the known
planets (see Table~\ref{prop}) are represented in each case by a
horizontal dotted line. These figures show the stability of the
planetary orbits remain secure for most values of the binary
eccentricity. The maximum binary eccentricities (where the planetary
semi-major axis lines intersect the eccentricity lines) are $e = 0.94
\pm 0.26$ and $e = 0.87 \pm 0.21$ for HD~2638 and HD~164509
respectively.

Given that the planets were discovered with the RV technique, it is
worth pausing to consider the effect of the stellar binary companions
on the planetary interpretation of the RV data. Using the stellar
parameters of the primary and secondary from Tables \ref{prop} and
\ref{isotable} respectively, along with the projected separations from
Table \ref{data}, we calculate the expected orbital periods and RV
semi-amplitudes for each system. For HD~2638, the minimum orbital
period is $\sim$110 years with a maximum RV semi-amplitude of
$\sim$2.4~km/s. For HD~164509, the minimum orbital period is $\sim$180
years with a maximum RV semi-amplitude of $\sim$1.7~km/s. As noted in
Section~\ref{properties}, the Kepler solution to the HD~164509 RVs
includes a linear trend, though the time baseline since discovery is
insufficient to charactize the nature of the trend. Assuming the
minimum separations above, the companions cannot be confused with the
planetary signals and thus have no effect on the planetary
interpretation of the RV data.


\section{Conclusion}
\label{con}

Determining the stellar architecture of planetary systems is an
on-going process, improving as the capability to detect faint stellar
companions increases. Stellar binarity can have a profound effect on
exoplanetary systems, both in terms of formation processes and
long-term orbital stability. Thus determining the binarity of known
exoplanet host stars is a critical step in the characterization of
those systems.

Here we have presented detections of stellar companions to two known
exoplanet host stars: HD~2638 and HD~164509. Though the stellar
companion to HD~2638 was previously detected by \citet{roberts15}, the
new data from DSSI will provide additional information of the
astrometry of the companion and the stellar properties, given that the
passbands used are particular to the DSSI camera. We have shown that
the detected companions have properties consistent with them both
being M dwarfs, and the isochrone analysis shows that they are both
likely to be gravitationally bound to the host stars. Fortunately, the
presence of the stellar companions do not pose serious orbital
stability problems for the known exoplanets, making the overall
architecture of the systems self-consistent. These planetary systems
represent additional interesting examples of planet formation and
evolution in the presence of multiple stars.


\section{Acknowledgments}

Based on observations obtained at the Gemini Observatory, which is
operated by the Association of Universities for Research in Astronomy,
Inc., under a cooperative agreement with the NSF on behalf of the
Gemini partnership: the National Science Foundation (United States),
the National Research Council (Canada), CONICYT (Chile), the
Australian Research Council (Australia), Minist\'{e}rio da
Ci\^{e}ncia, Tecnologia e Inova\c{c}\~{a}o (Brazil) and Ministerio de
Ciencia, Tecnolog\'{i}a e Innovaci\'{o}n Productiva (Argentina). This
research has made use of the NASA Exoplanet Archive, which is operated
by the California Institute of Technology, under contract with the
National Aeronautics and Space Administration under the Exoplanet
Exploration Program. The results reported herein benefited from
collaborations and/or information exchange within NASA's Nexus for
Exoplanet System Science (NExSS) research coordination network
sponsored by NASA's Science Mission Directorate.


\end{document}